# A Logical Interpretation of a Delayed-Choice, Quantum Eraser Experiment

William Sellers, Wright State University*
[william.sellers@wright.edu]

*ABSTRACT: This paper shows how one can construe the experimental results in a way that does not involve effects that precede their causes.*

In their paper "Quantum entanglement: from Popper's experiment to quantum erase" [1] Y. Shih and Y.-H. Kim describe the following experiment [see Fig. 1]:

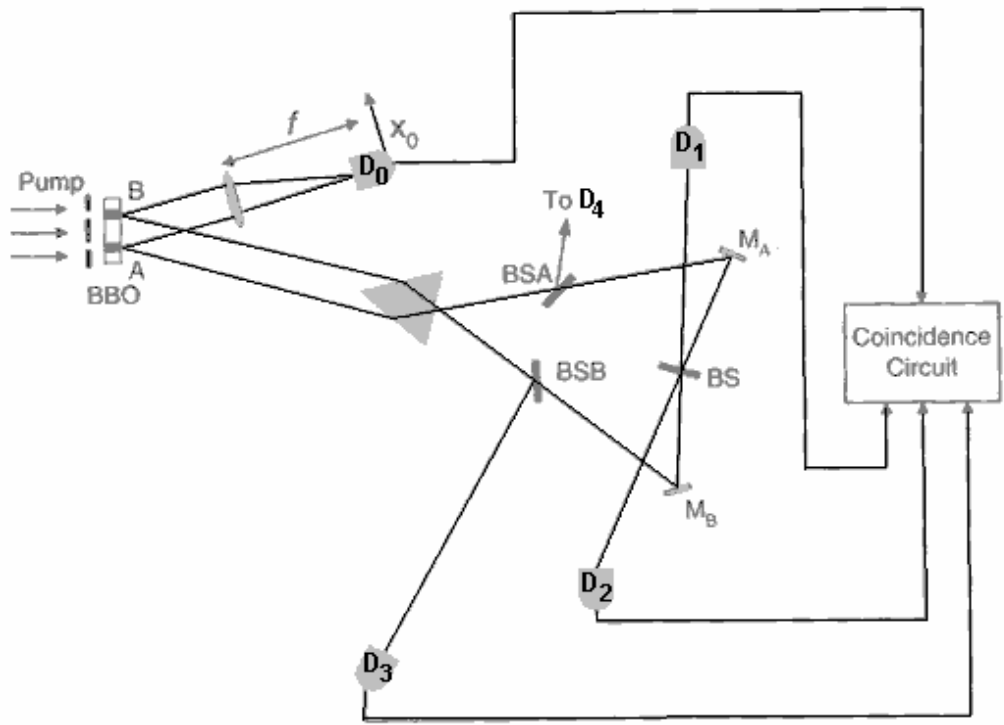

Fig. 1

[a] laser beam is divided by a double-slit and directed onto a type-II phase matching nonlinear optical crystal BBO…at regions A and B. A pair of…orthogonally polarized signal-idler photon is generated either from A or B region….A Glen-Thompson prism is used to split the…signal and idler. The signal photon (photon 1, coming either from A or B) propagates through a lens to

trigger detector $D_0$,…The idler photon (photon 2) is sent to an interferometer with equal-path optical arms. (pp. 364-5)

BS, BSA, and BSB in the figure are 50-50 beam-splitters, MA and MB are reflecting mirrors, and there are detectors at $D_1$, $D_2$, $D_3$, and (not shown) $D_4$. Photon 1 and photon 2 are 'entangled,' with photon 1 being detected at $D_0$ and photon 2 being detected at $D_1$, $D_2$, $D_3$, or $D_4$. For the sake of convenience, let us call a signal photon that travels through slit A, photon $1_A$, and if it travels through slit B, photon $1_B$. Similarly we will have either idler photon $2_A$ or idler photon $2_B$.

The experiment is set up so that if the detection is at:

$D_1$: it could have been triggered by either photon $2_A$ or photon $2_B$

$D_2$: it could have been triggered by either photon $2_A$ or photon $2_B$

$D_3$: it could only have been triggered by photon $2_B$

$D_4$: it could only have been triggered by photon $2_A$

Thus the last two detections would give the experimenter "which path" information about photon 1 but the first two would not.

When the experiment is run, it turns out that coincidences between $D_1$ and $D_0$ and those between $D_2$ and $D_0$ both show interference but that coincidences between $D_3$ and $D_0$ do not. (No data for $D_4$ were published but the authors say that the results were similar to those for $D_3$.)

> The experiment is designed in such a way that $L_0$, the optical distance between atoms A, B and detector $D_0$, is much shorter than $L_A$ ($L_B$) which is the optical distance between atoms A, B and the beam splitter BSA (BSB). Thus after $D_0$ is triggered by photon 1, photon 2 is still on its way to the first beam splitter and does not 'know' 'where' to go yet. (pp. 363-4)

We have here an apparent instance of what Einstein famously called "spooky action-at-a-distance," possibly even an example of a photon going back in time and changing the past. And yet the experiment is entirely consistent with, and predicted by, standard quantum mechanical calculations. Is there an interpretation of the results that does not fly in the face of common sense?

The interpretation to be offered here complements one offered by Srikanth [2] with respect to Wheeler's "delayed-choice" thought experiment. In the Wheeler setup, the photons pass through both slits (as Srikanth's analysis shows), and so in the region of the slits there are no which-path constraints. But the detection apparatus either does or does not filter out amplitude information for one of the slits. If it does not, no path constraints are imposed and interference appears. If it does, then a path constraint is imposed *by the detector* and no interference is seen.

In the experiment under review, on the other hand, the entangled photons are generated in such a way that at BBO there are which-path constraints. Note in Fig. 1 that *all* the photons arriving at BSB (or BSA) are correlated with photons that have passed through only one slit. No "choice" is made at BSB (or BSA)—half are simply reflected to $D_3$ (or $D_4$) and the other half pass through the beam splitter. It doesn't matter which idler photons are reflected and which pass through: no interference can be detected because the correlated signal photons can have only one path.

Similarly, all the idler photons arriving at $D_1$ (or $D_2$) are correlated with signal photons that do not have which-path constraints. It is this lack of constraint that allows for the interference to manifest itself. What the detectors $D_1$ and $D_2$ do is to *combine* amplitude information for both slits. Detectors $D_3$ and $D_4$ only receive amplitude information from one slit.

It is fairly easy to see how the detectors in the Wheeler 'delayed-choice' thought experiment filter out amplitude information (the telescope is oriented in such a way that light from only one slit reaches the eye-piece even though light from both slits strikes the objective lens), but perhaps not as obvious how the detectors $D_1$ and $D_2$ in the current experiment combine information. The combination is achieved through the coincidence circuit.

Instead of having $D_0$ move along the x axis, let us imagine that we have a number of detectors located along this axis and that each detector is connected to a stopwatch. [3] When there is only one slit, the situation will be as shown in Fig. 2:

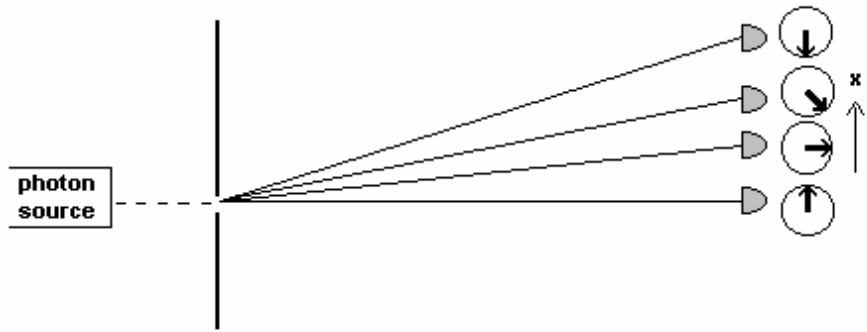

Fig. 2

As we move further up the x-axis, the photons take longer to reach the screen. Similarly, when we have a double-slit [see Fig. 3],

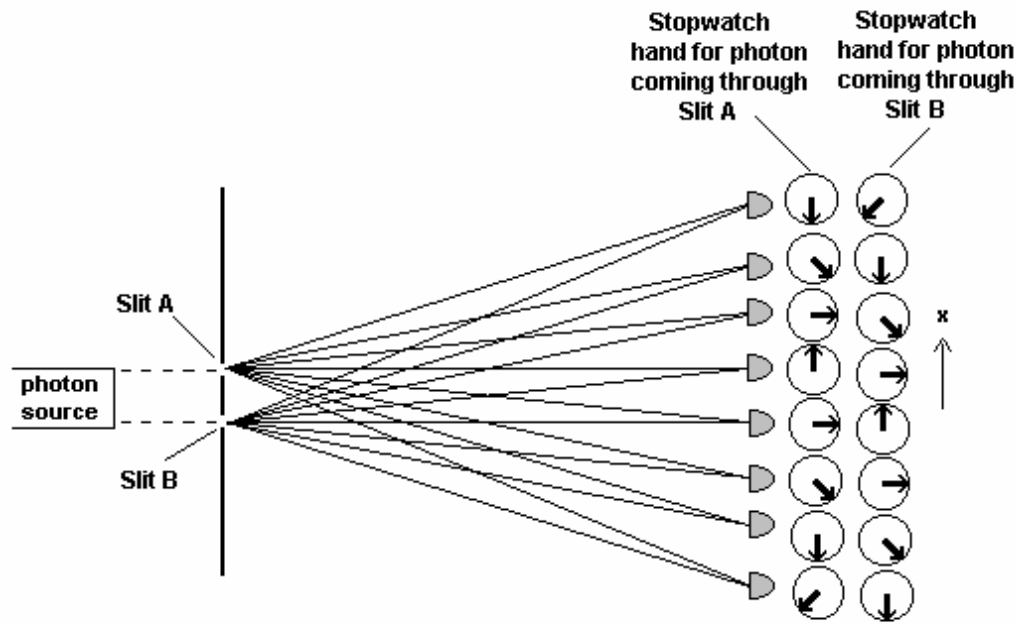

Fig. 3

the photons, with one exception, never arrive at the screen at the same time. (The exception is the spot directly in front of the middle of the two slits.) Figs. 2 and 3 are, of course, not drawn to scale; the stopwatch hands are not intended to show what their actual positions would be. Nevertheless, the principal is the same—namely, that there is a certain interval of time ($\Delta t$) that separates each click at a particular location along the x-axis. In Fig. 3 this interval ranges from a 45º to a 90º difference in the position of the two stopwatch hands.

➢ If we were using a stopwatch that was accurate to a range smaller than $\Delta t$, and we used single photons whose time of emission from the source was also timed with at least as accurate a stopwatch, then we would have "which path" information for each photon and the photons coming from either slit A or slit B would appear to arrive distinctly. Presumably the detectors we placed along the x-axis in Fig. 3 would then not detect interference.

➢ If we were using a stopwatch that was accurate only to a range *larger* than $\Delta t$, and if we used single photons (regardless of how accurately their times of emission from the source were measured), then the photons coming from either slit A or slit B would appear to arrive simultaneously and we would *not* have "which path" information for each photon. Presumably the detectors we placed along the x-axis in Fig. 3 would then detect interference.

Looking back at Fig. 1, we see that in order for detectors $D_1$ and $D_2$ to be described as "not providing which-path information," photons coming from either slit A or slit B must be considered to be arriving at the detector at the same time. Since we know that the times are not the same when measured to an accuracy within $\Delta t$, this must mean that we have set the coincidence interval for the coincidence circuit (i.e., the period of time within which two signals will be regarded as being coincident) greater than $\Delta t$. Thus the amplitude information for the photons arriving at $D_0$ will be combined and interference will appear.

*I should like to thank Leno Pedrotti, University of Dayton, for his comments on an earlier draft of this paper.